# Did Maxwell dream of electrical bacteria?

Eleonora Alfinito [1,*], Maura Cesaria [2] and Matteo Beccaria [3,4]

[1] Department of Mathematics and Physics 'Ennio De Giorgi', University of Salento, Italy; eleonora.alfinito@unisalento.it
[2] Department of Mathematics and Physics 'Ennio De Giorgi', University of Salento, Italy; maura.cesaria@unisalento.it
[3] Department of Mathematics and Physics 'Ennio De Giorgi', University of Salento, Italy; matteo.beccaria@unisalento.it
[4] National Institute for Nuclear Physics, INFN, Italy; matteo.beccaria@le.infn.it

* Correspondence eleonora.alfinito@unisalento.it

**Abstract:** We propose a model for bacterial Quorum Sensing based on an auxiliary electrostatic-like interaction originating from a fictitious electrical charge that represents bacteria activity. A cooperative mechanism for charge/activity exchange is introduced to implement chemotaxis and replication. The bacteria system is thus represented by means of a complex resistor network where link resistances take into account the allowed activity-flow among individuals. By explicit spatial stochastic simulations, we show that the model exhibits different quasi-realistic behaviors from colony formation to biofilm aggregation. The electrical signal associated with Quorum Sensing is analyzed in space and time and provides useful information about the colony dynamics. In particular, we analyze the transition between the planktonic and the colony phases as the intensity of Quorum Sensing is varied.

## 1. Introduction

When, in the 60s of the XIX century, J.C. Maxwell wrote the milestone equations that take its name, he probably did not have in mind their applications to living matter. On the other hand, from the perspective of physical interactions, living matter is not different from inorganic matter because only gravitational and electromagnetic forces are relevant on a mesoscopic scale (1μm-1nm). Furthermore, electrical interactions govern many chemical and biochemical phenomena. Only when the length scale reduces to the atomic size, quanto-mechanical effects become relevant also in the biological realm; for example, electrical transport, which is at the basis of many fundamental biochemical phenomena, like photosynthesis, is basically due to the quantum tunneling effect [1].

With a size of about 1μm, cells are the smallest living being able to perform the basic tasks of life, i.e. they are able to reproduce, get food and perform complex actions. Often, they are part of a more complex system, but some of them (bacteria, algae etc.) may survive and function alone. Bacteria are found both in the planktonic state (single cells free of moving in a culture medium) and in colonies. The transition from one state to the other may be driven by the opportunity to improve their chances of survival [2]. Association in a colony often contemplates the formation of a biofilm, i.e. a polymeric matrix in which bacteria are encased. It is a utilitarian behavior that requires cooperation and is observed both in liquids, where swimming cells aggregates in flocks/suspensions [3] and biofilms formed onto solid surfaces [4] . As a matter of fact, when bacteria come and work together, they become stronger against the attacks of enemies, while, as a drawback, they have to compete for food. The difference with the planktonic state is not only qualitative but also quantitative, since bacteria in colonies show a higher gene expression [4]. On the other side, cooperative effects have been also observed in swimming bacteria, specifically, the speed of a coordinate set of bacteria is until five-fold larger than that of the individual cell and it grows for increasing density of the set. [3, 5]. The mechanisms for the development of this collaborative strategy are commonly known as Quorum Sensing (QS) [6]. The detailed biochemical specifically depends on the bacteria phenotype, although some general features may be highlighted [7]. They rely on the synthesis of autoinducers (AIs), their release in the cell environment, and their capture by specific receptors present on different cells, thus acting on the regulation of gene expression [8]. When the cell concentration overcomes an appropriate threshold, the associated AI concentration becomes sufficient to trigger a collective change in gene expression which produces a strong signal in terms of bioluminescence, antibiotic production, cell duplication, biofilm formation, etc. [7, 8].

Due to the interest and fascination of the topic, several models have been proposed to describe specific processes like for instance biofilm development [9, 10, 11], swimming bacteria [3], bioluminescence activation [12], antibiotic production [13].

Just as the topic is wide and complex, so are the models. They include analytical investigations ([9,12], stochastic approaches [14,15], and agent-based models [16, 11]. This wide variety of models addresses specific problems related to the biomass formation [9], autocatalytic cycles underlying the growth laws [17], as well as migration and colony growth [14].

In this paper, we propose a stochastic model that borrows from electrostatics the key element of an interaction among far (charged) elements, which is a form of communication, and uses this language to mimic the bacteria evolution, migration and colony formation.

In more details, the model describes the evolution of a bunch of initial seeds, i.e. bacteria, randomly distributed on a regular graph and equipped with an initial kind of 'internal energy' [14], herein called activity. Activity reflects the level of gained genic regulation and plays the role of an electric charge, hence allowing each bacterium to electrically interact with other alive bacteria. A network of resistance channels connect live bacteria (nodes) and the exchange of AIs among them is described by a reduction of the resistance of the associated channel. Each bacterium establishes a sub-network of links with other alive bacteria, thus mimicking the AI exchange mechanism. The activity of a node grows accordingly to the number of received AIs and therefore it depends on the number of alive bacteria. When the activity is low, each bacterium may migrate toward sites with lower electrostatic potential, which means regions in which it has higher chances to increase its activity and replicate. Activity can grow until a maximal value, after that the bacterium dyes giving back its energy to the system. This dynamics aims to reproduce the life cycle in which an individual consumes food and occupies space until its death, after leaving space and food to a novel individual [14]. The current flowing inside the network is tracked and represents the degree of global activation, which depends on a few model- parameters representing the effectiveness of AIs exchange, the initial number of seeds, and the cooperation among bacteria.

Furthermore, the associated QS signal is produced in terms of the charge flux inside the network and may be used as a colony wellness indicator.

Our results show that, although the proposed model is based on rather general interaction principles, it works in the prediction of QS phenomena by combining two main parameters (activity and activity efficiency rate).

**Materials and Methods**

A regular resistance network is used to represent the space in which the colony may develop and the AIs travel. Nodes represent the empty/filled positions of the bacteria and are arranged on a regular rectangular $L_x \times L_y$ grid.

Each node may take different levels of activation, denoted by Q. In this context, activation describes the ability of bacteria to perform specific activities like gene regulation. The activation of node n is denoted by Q(n) and is an integer quantity which is positive for activated sites, zero for empty sites, and -1 for blocked sites. The quantity Q is expected to grow in fair environmental conditions (for instance, no predators, free food, physiological values of temperature and pH) and decreases otherwise, for example, resulting in sporulation, death, or biofilm formation. The higher the activity, the higher the ability to replicate, or infect other organisms, or possibly produce light. On the reverse, the more intensive QS, the higher is the activity [18]. This interconnected *modus operandi* is here reproduced, at least qualitatively, by dealing with Q as if it was an electric charge, and allowing different nodes to mutually interact by means of an electric-like Coulomb interaction. In our model, the probability of interaction becomes higher at increasing level of activation. In the present investigation, we analyze the effects of QS on the colony development and, therefore, higher activity will imply a higher probability to reproduce [19]. Each node configuration will be associated with a corresponding potential distribution that regulates the colony evolution. Quorum Sensing emerges as a consequence of the non-trivial dynamics and is monitored by changes of the electrical features of the network which, as the AIs progressively diffuse, reduces its resistance. In such a way, it is possible to follow both the colony development and the progression of QS.

The detailed stochastic process describing the colony evolution goes through several steps in sequence. In all cases, an initial randomized distribution of nodes -seeds- is setup by assigning the value Q

=1 to a fraction $f_0$ of the nodes in the grid. When this number is increased, the probability that the system evolves in migration or colonization increases. The evolution steps are illustrated in details below.

1) In the first step, the potential V of each node and the energy of the whole network is computed. For the $l$-th node, the potential $V(l)$, and its energy, $\varepsilon(l)$, are, respectively:

$$V(l) = \sum_{j \neq l}^{N} \frac{Q(j)}{Dist(j,l)}, \qquad \varepsilon(l) = Q(l) \sum_{j \neq l}^{N} \frac{Q(j)}{Dist(j,l)}, \qquad (1)$$

where $N = L_x \times L_y$ is the network size and $Dist(j,l)$ is the Euclidean distance between the two nodes $l$ and $j$.

The energy of the network is computed by evaluating the Coulomb interaction

$$Energy = \frac{1}{2} \sum_{\substack{i,j=1 \\ (i \neq j)}}^{N} \frac{Q(i)Q(j)}{Dist(i,j)} \ . \qquad (2)$$

Both the potentials and the energy change with the iteration step (time).

2) Nodes establish contacts with nodes having a lower potential. Also, the probability of establishing a contact is higher for nodes with mutually closer energy values. This implements the idea that nodes with higher activity sends AIs toward nodes with lower activity. Thus, for any pair of nodes with labels n and m, we first sort them in order to have ε(n)>ε(m) and then a connection n-m is activated with probability

$$p(n,m) = \min(1, \exp(-\beta \ \Delta E_{n,m})), \qquad \Delta E_{n,m} = \frac{\varepsilon(n) - \varepsilon(m)}{Energy} \qquad (3)$$

which is chosen in order to favor the nodes closest in normalized energy. The choice of a Boltzmann-like linking probability is somewhat arbitrary. Generally speaking, it can be replaced by similar functions without changing qualitatively the outcome of the evolution. In our simulations, we will take β=1.

3) A pair of extended ideal electrical contacts is put at the ends of the network [20,21]. When a contact is established between nodes n and m, we assign to the link a resistance in the interval [$r_{min}$, $r_{max}$] according to the formula:

$$res(n,m) = Dist(n,m)[r_{max}(1 - f(n,m)) + r_{min} f(n,m)], \qquad (4)$$

where the interpolating functions $f(n,m)$ is taken to have a Hill-like shape [10,12]:

$$f(n,m) = \frac{Q(n)Q(m)}{g + Q(n)Q(m)} \ . \qquad (5)$$

The parameter g controls the steepness of the interpolation and, in our simulations, will be fixed at the value *g*=0.01. Other choices for *f(n,m)* are with the constraint of being unity for large activation product *Q(n)Q(m)*. The Hill form in (5) is just the simplest possibility. The change in the resistance represents the QS signal. As a matter of facts, the resistance reduction means that an electrical current is flowing between a couple of nodes, thus accounting for the transfer of information among those nodes (AI exchange) [20,21]. The opening of conduction channels, due to the interpolation effect of the $f$ function (Equation 5), becomes even more sharp at increasing value of activity and mimics the autocatalytic effects observed in colony growth [17]. The role of $f$ is that of implementing the cooperative action of different AIs and assuring a smooth evolution of the network with iteration time too.

4) For each node n, the number of activations produces a score which is added to the node activity *Q(n)*. A certain amount of charge is distributed among the nodes that are connected. In particular:

$$Q(n) \rightarrow Q(n) + floor\left(\frac{\sigma * links(n)}{N}\right), \qquad (6)$$

where $1 < \sigma < N$ is a real number whose value determines the efficiency of activation, and *links(n)* is the number of nodes connected to the *n*-th node.

5) In this step, for each (parent) node we consider migration/duplication transitions. First, we choose one empty node out of the 8 nearest neighbors. This choice, as highlighted in the remarkable paper [14], is not purely random. It is driven by utilitarian reasons like the reach of regions with higher amount of food or different bacterium concentration. The selected node is called the target node. The choice is done by first sorting the neighbors in order of increasing potential. Then, the k-th node in the list is selected with probability

$$p(k) = \frac{(k-1)!}{9^{k-1}}\left(1 - \frac{k}{9}\right), \qquad k = 1, \dots, 8. \tag{7}$$

This formula corresponds to choose the minimum potential node (k=1) with probability 1-1/9 and otherwise, with probability 1/9, choose the second (k=2) with probability 1-2/9 and so on.

If the parent node has the minimum nonzero value Q=1 it will migrate to the target node that inherits Q=1 while the parent node is set to Q=0.

If instead the parent node has Q≥ 2 a daughter-daughter reproduction is implemented and the parent node gives half of its charge to the target. The daughter-daughter >-reproduction is almost similar to the binary fission, in which the original cell splits into two equal parts, and is the most credited framework for bacteria replication [22].

6) When Q(n) reaches the assigned maximum value $Q_{max}$, we consider again two possible rules. In the first rule(**DYING**) the node dies, i.e. Q(n) =0. In the second one (**STATIC**) it evolves toward a static form with Q(n) = -1 (spore-biofilm). The value of the maximum allowed charge $Q_{max}$ can be tuned and when it is increased, the evolution time becomes longer. In all our simulations it will be fixed at the value $Q_{max}$=80.

In the **DYING** scheme, the final exit of the bacterium evolution is death and its activity returns to the network and can be reused and the bacterium replaced. The network reaches a stationary state with a final mean energy and some nodes which are continuously reborn. This condition mimics the formation of a swimming colony (flocks) [3] which behaves in a cooperative manner. In the **STATIC** scheme, the node becomes inactive and cannot be substituted. In this case, its activity remains trapped and the network reaches a quasi-static state with few alive nodes. This condition mimics the formation of a stationary colony (biofilm) in which few cells or microcolonies are encased in a polymeric matrix.

Notice that, in this version, interactions among faraway sites describe the exchange of signaling AIs among bacteria, therefore, they happen only between active nodes. Steps (1-2), (4-6) pertain to the colony dynamics (AIs diffusion, gene regulation of reproduction, colony formation) while step (3) describes QS signal.

In summary, the free parameters of our simulations are listed in the following Table I (in the third column we report the fixed values used in the presented simulations)

**Table 1.** Model parameters. We briefly recall their meaning and the values adopted in simulations.

| $L_x$, $L_y$ | Dimensions of the rectangular grid | variable |
|---|---|---|
| $f_0$ | Initial fraction of occupied nodes | variable |
| $\beta$ | Parameter entering the linking probability | 1 |
| $r_{max}$, $r_{min}$ | Resistance values entering the link resistance formula | $r_{max}$=1000, $r_{min}$=1 |
| g | Parameter in the Hill-like function, controlling the resistance interpolation | 0.01 |
| $\sigma$ | Parameter controlling the activation efficiency | variable |
| $Q_{max}$ | Maximum value of the activity triggering death or biofilm formation | 80 |

Also, due to the two options in both steps (5) and (6), there are actually 2 possible simulation rules (submodels) corresponding to the choices **DYING/STATIC**.

### 3. Results

Simulations were performed using the parameters reported in Table 1, specifically, a. two different network geometries, rectangular and squared; b. different values of $f_0$, to produce simple migration or colony formation; c. two possible ending scenarios which, roughly speaking, correspond to the formation of colonies or biofilms.

*3.1. DYING model*

A colony is formed only when an appropriate combination of the initial density of seeds, $f_0$, and the activity efficiency rate, $\sigma$, occurs. To exemplify this concept, in Figure 1 the mean value of active nodes vs. $\sigma$ is reported. The analyzed range of $\sigma$ is 10-30 and the most noisy-data come from the range 19-22, showing a rather sharp transition. In other terms, while at the ends of the $\sigma$ range each realization produces migration/ colony with probability close to 1, in the intermediate region the same value of $\sigma$ may result in both migration or colony formation.

Averaging is made over 11 realizations, dropping the first 30 iterations, using $f_0$=0.05 and the rectangular geometry 200x10.

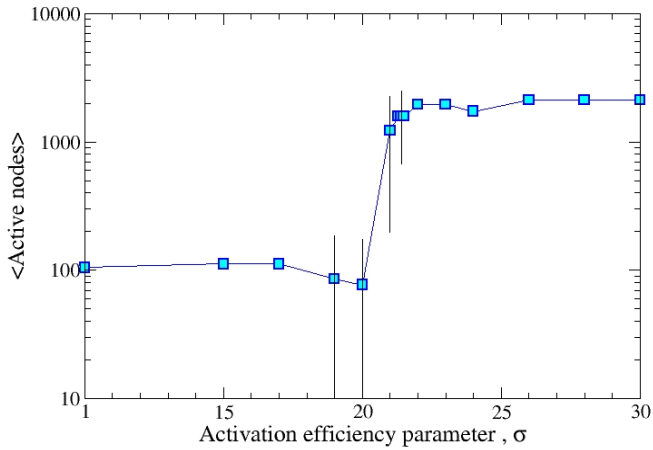

**Figure 1**. Transition from migration to colony formation regime. The transition is represented by the mean value of the active nodes vs. the activity efficiency rate, $\sigma$. Each node corresponds to the mean calculated over 11 realizations, using $f_0$ = 0.05 on a 200x10 network. Line is a guide for the eye. The error bars are reported when significant. The transition from migration (few active nodes) to colony formation (200x10 active nodes) is quite sharp in the range of $\sigma$= 19-22.

In Figures 2a, b two different kinds of evolution are illustrated, obtained with a quite large initial fraction of seeds (5% of the total number of nodes, N) and a small value of $\sigma$ (21.5) in a 200x10 network. Two different realizations are reported: one outcomes in a simple migration, Figure 1a, the other evolves in a colony. In Figure 2c we present the electrical current inside the network as a monitor of the QS signal. Pure migration produces a noisy low signal. When concentration reaches a value large enough to start colony formation, the signal become larger and much less noisy. Simulations were performed assuming the condition DYING, see section Methods, and this is responsible for the fluctuating asymptotic behavior of the QS signal. In other terms, the simulated electrical current is a good indicator of the colony status and of the QS occurrence.

The geometry of the network may also affect the outcome of the simulation. A complete investigation, along the lines of what was made e.g. in [23,24] is left for future work. Nevertheless, as a preliminary comparison, we present in Figure 3, simulations performed on a 50x50 square network.

In Figure 3a two different evolutions are analyzed, one with $f_0$=0.05 and $\sigma$=22, the other with $f_0$=0.015 and $\sigma$=100. The second realization starts with less seeds, but, due to the higher activity growth rate, $\sigma$, in few iterations (at iteration 10) overcomes the first realization. Also, the dying and recovery process is faster and therefore the mean value of active nodes is smaller. This result agrees with the smaller asymptotic value of the QS signal (Figure 3b). The faster onset of the colony formation regime is also detectable from the QS signal.

Figure 4 illustrates the same colony evolution as in Figure 3a (black). Here, the color scale displays the values of Q on the network. After a first phase in which the seeds migrate in order to increase their density, Q also becomes to grow, the highest values remain in the central region, in qualitative agreement with data coming from microscopic analysis literature [25].

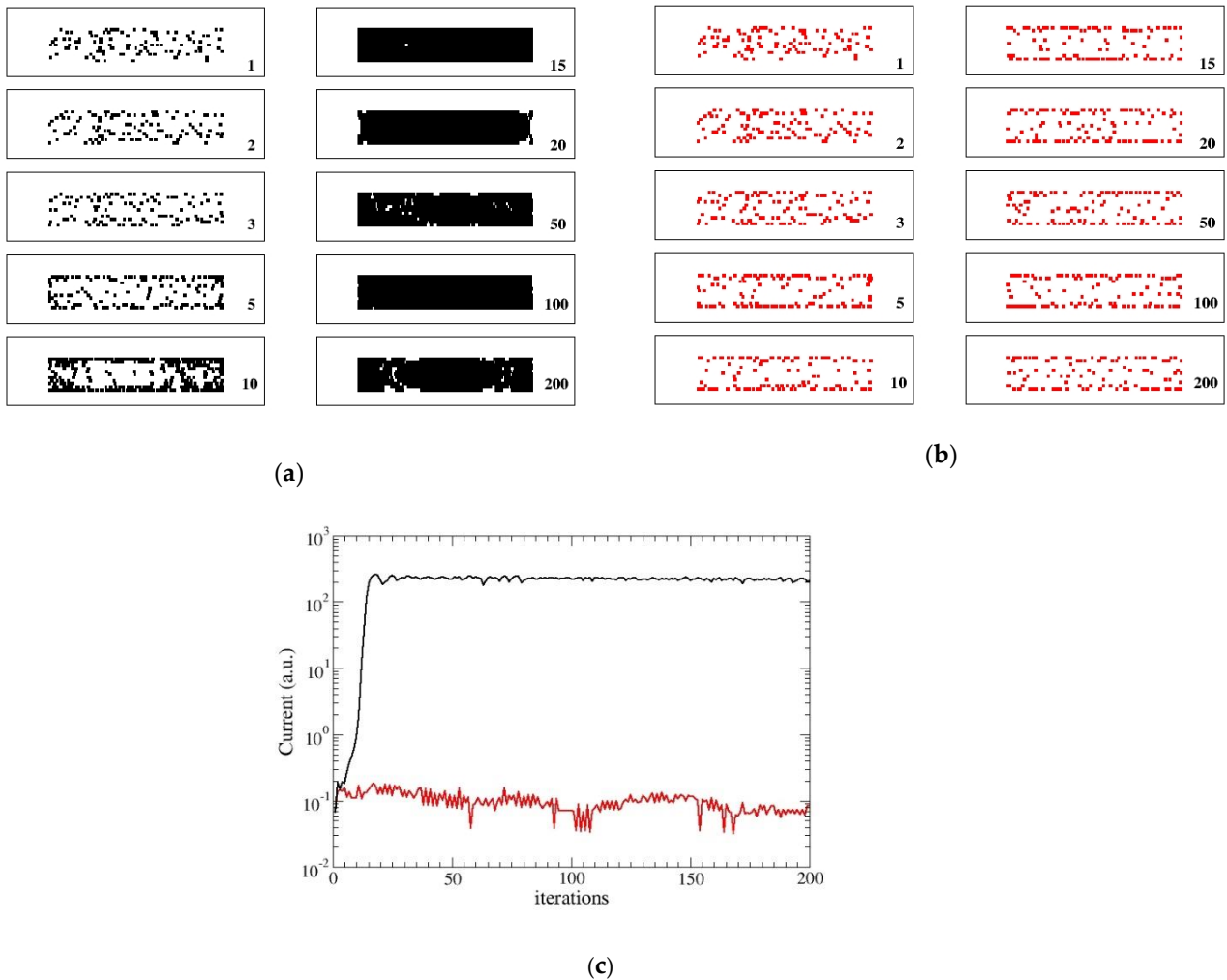

(**a**)

(**b**)

(**c**)

**Figure 2**. Colony formation in a rectangular lattice. The initial percentage of seeds is 5% of the maximal space occupancy. In the transition region (here σ=21.5) the system can exit into a simple migration (**a**) or a colony formation (**b**). Each screenshot of (**a**) and (**b**) is obtained at a different iteration time, from 1 to 200. (**a**) Screenshots of the migration evolution of an initial seed distribution. In the first three screenshots seeds (black pixels) simply migrate. (**b**) Colony formation with same conditions as in Figure 1a. (**c**) Monitoring of the QS signal. The electrical current of the network mirrors the QS for the development of a colony (black), and a migration (red).

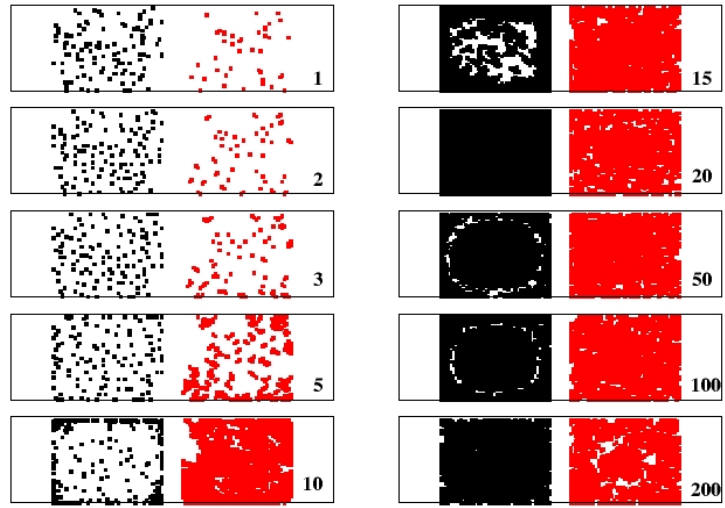

(a)

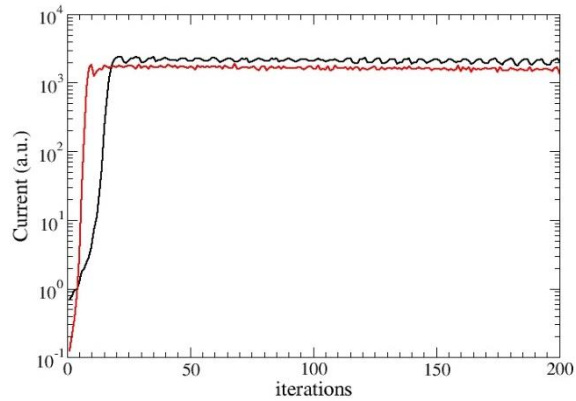

(b)

**Figure 3**. Colony formation in a 50x50 square lattice. Two different kinds of evolution are sketched. In black color: a quite large initial number of seed (5% of total amount) with a small rate of reproduction (σ=22). In red color: a smaller initial number of seeds (1.5 % of total amount) and a large rate of reproduction (σ=100). (**a**) Screenshots of colony formation taken at different evolution times from 1 to 200. (**b**) QS signal. The electrical current of the network reflects the QS for the development of a colony. The faster reproduction rate in the setup with higher value of σ (red) with respect the that with lower value of σ (black) is well reproduced.

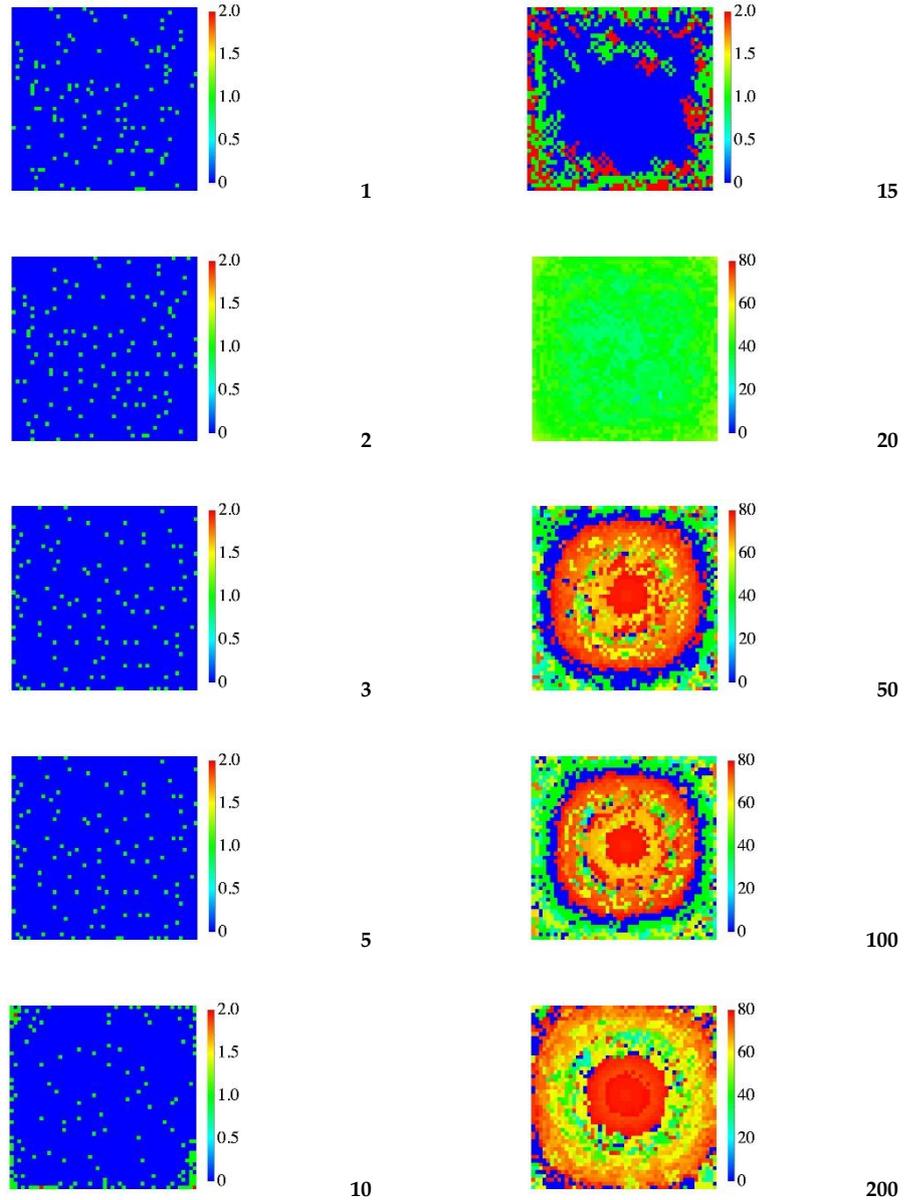

**Figure 4**. Colony formation in a square lattice. The color scale represents the Q values. The screenshots refer to a 50x50 network, initial amount of seeds, $f_0$=0.05, activity growth rate, $\sigma$=22. This picture corresponds to Figure 3a (black dots).

*3.2. STATIC model*

The STATIC scenario has also been investigated both in rectangular and square geometries, cf. Figure 5. In this case, we remind that when a bacterium reaches $Q_{max}$, it evolves into a biofilm component. In such a way, it does not move anymore and does not evolve further (no reproduction, no change of score, no connection with other nodes). Biofilm formation requires a high cost in terms of energy [2] and this should result in the total emptying of the nodes (Q=-1) (and corresponding nodes stop interacting with the others) The QS signal reduces rapidly to a quite small value. For a sufficiently long time, the colony signal becomes stationary without sizable noise. The formation of microcolonies, i.e. small clusters of alive bacteria, is observed in both the network configurations. In these cases, bacteria do not die because they do not reach the maximum value of activity. Figure 5c shows the associated QS signal which accounts for the different times at which the colony is fully formed (the maximum of the curve) as well as for the different amount of signal (larger in the square lattice which contains more nodes).

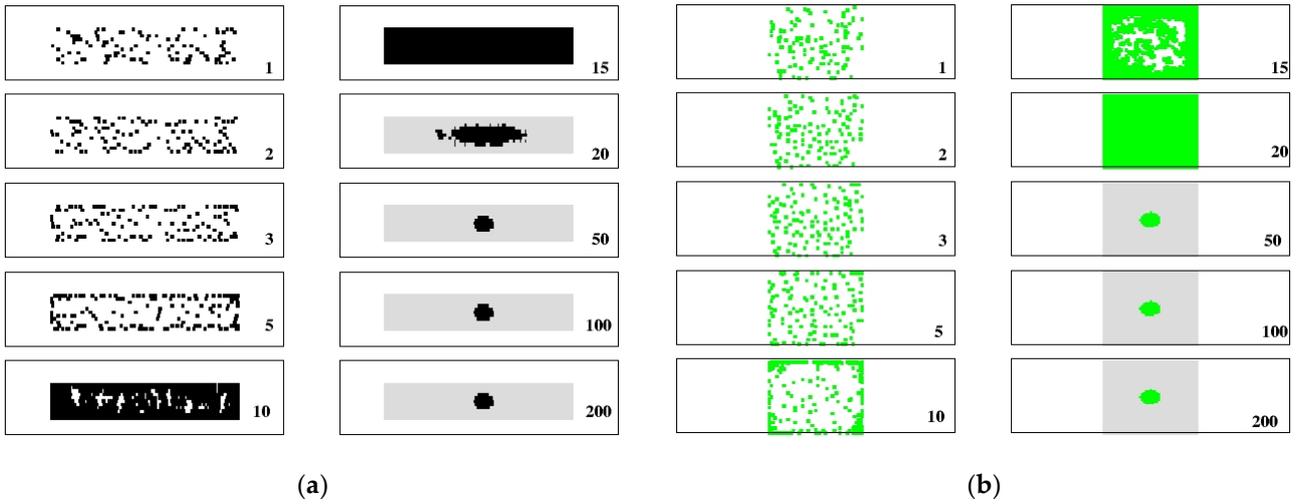

**Figure 5.** Biofilm formation. Two different geometries are sketched: (**a**) 200x10 and (**b**) 50x50. In both cases, the same evolution conditions are applied, i.e. equal initial number of seed (5% of total amount) with a small rate of reproduction ($\sigma$ =22), in black and STATIC scenario . (**a**) Screenshots of the colony formation taken at different evolution time from 1 to 200. Alive bacteria (black) and biofilm (gray) co-exist. (**b**) Screenshots of the colony formation taken at different evolution time from 1 to 200. Alive bacteria (green) and biofilm (gray) co-exist. (**c**) QS signal. The electrical current of the network mirrors the QS for the development of a colony (black for rectangular network, green for square network).

## 4. Discussion

Quorum sensing represents the ability of bacteria to coordinate their activities in order to improve performance when working together instead of being isolated. This feature is observed both in free moving cells [3] and in stationary cells which aggregate on a substrate [2,6-8]. Despite its underlying and fundamental role in many bacteria features, QS is itself not measurable except in its global manifestations swimming velocity, replication, bioluminescence, etc.).

In our approach, the evolution of a bunch of seeds (bacteria) is described in terms of abstract auxiliary electrostatic-like interactions. Each cell is endowed with a quantity Q called *activity* that measures the ability to replicate or, in more general terms, to perform the specific features of its phenotype. The chosen evolution laws can drive toward a scenario in which the cells reproduce, reaching a quasi-stationary state. In this state the number of active sites fluctuates around a mean value while activity concentrates inside the colony. The spatial structure of the system is that of a complex resistance network [1,20,21] where nodes are the bacteria and links are determined dynamically by their interactions. In the absence of interaction between a pair of nodes, the resistance of the associated link is maximal, representing a closed channel. When interaction is relevant. the link resistance becomes even more smaller and this effect is enhanced when the activity of the connected nodes becomes larger. In this description, the QS signal is represented by the electrical current flowing in the network and appears like a noisy low intensity signal when colony formation does not start; otherwise, its intensity grows sharply and the shape agrees with the kind of evolution (colony /biofilm formation).

In the present investigation, QS is only related to the colony formation, but more generally it may be responsible for other collective physical signals, as bioluminescence. As a matter of facts, although qualitatively, bioluminescent signals [26] have certain similarities to the QS signal that we observe in our simulations during biofilm formation. The analysis of the statistical properties of the observed QS signal will be carried out in future developments, in agreement with previous studies [23,24].

The simple model presented here has many parameters and we just begun to explore their interrelated roles by some restricting choices. Other regions of this still uncharted parameter space may correspond to more complex bacteria features. Natural extensions are easily included with minor modifications, like different kind of replications, the presence of multiple AIs, or bacteria species and will be made in accordance with experimental data in the process of being acquired.

### 5. Conclusions

The phenomenology of bacteria behavior is wide and complex and, surely, a simple (and abstract) model like the proposed one cannot fully capture it. In particular, there are still important missing features with respect to real systems (kind of replication, different types of AI, and so on). Nevertheless, as we illustrated by explicit simulations, several complex behaviors mimicking real phenomena emerge and justify the formulation of a bacterial toy model. Specifically, the proposed model is as simple as possible being based only on rather general interaction principles and flexible enough to be adapted to specific individuals. The aim of this paper has been highlighting how a simple (auxiliary) electric interaction gives rise to a non-trivial complex dynamics. This remark could be the starting point for more complex models in the same spirit with the aim of coming closer to real data.

One important open issue concerns the possibility of upgrading the fictitious electrostatic interaction to a physical one. This stems from the simple remark that electrical bacteria do exist. Many electroactive microorganisms are known [27,28] which are able to exchange electrons with their environment and that, as non-electric bacteria, live both in planktonic and biofilm assays [27,28]. Also, in this kind of microorganisms QS has a regulatory function, enhancing their electrical activity [28]. Thus, it is tempting to interpret the proposed complex electric network as some sort of effective description of real interactions in this class of microorganisms. According to this admittedly optimistic attitude, we can conclude that, after all, there is a chance that "Maxwell may have dreamt of electrical bacteria". [1]

**Author Contributions:** Conceptualization, E.A; methodology, E.A.,M.B.; investigation, M.C; writing, E.A., M.B., M. C.; funding acquisition, M.C. All authors have read and agreed to the published version of the manuscript.

**Funding:** M.C. was supported by the BANDO POR PUGLIA FESR-FSE 2014 / 2020 - Research for Innovation (REFIN)-Regione Puglia- Proposal Number: 012C1187.

**Data Availability Statement:** Not applicable.

**Conflicts of Interest:** The authors declare no conflict of interest.

## References

1. Alfinito, E.; Cataldo, R.; Reggiani, L. A pH-based bio-rheostat: A proof-of-concept. *APL*, **2022**, *120(1)*, 013701.
2. Flemming, H. C.; Wingender, J.; Szewzyk, U.; Steinberg, P.; Rice, S. A.; Kjelleberg, S. Biofilms: an emergent form of bacterial life. Nature Reviews Microbiology, **2016**, *14(9)*, 563-575.
3. Sokolov, A.; Aranson, I. S.; Kessler, J. O.; Goldstein, R. E. (2007). Concentration dependence of the collective dynamics of swimming bacteria. *PRL*, **2007**, *98(15)*, 158102.

---

[1] Freely adapted from Philip K. Dick's masterpiece "Do Android dream of electric sheep" (1968).

4. Resch, A.; Rosenstein, R.; Nerz, C.; Götz, F. Differential gene expression profiling of Staphylococcus aureus cultivated under biofilm and planktonic conditions. *Appl. Environment. Microbiol.*, **2005**, *71(5)*, 2663-2676.
5. Sabass, B.; Koch, M. D.; Liu, G.; Stone, H. A.; Shaevitz, J. W. Force generation by groups of migrating bacteria. *PNAS*, **2017**, *114(28)*, 7266-7271.
6. Fuqua, W. C.; Winans, S. C.; Greenberg, E. P. Quorum sensing in bacteria: the LuxR-LuxI family of cell density-responsive transcriptional regulators. *J. Bacterial.*, **1994**, *176(2)*, 269-275.
7. Ng, W. L.; Bassler, B. L. Bacterial quorum-sensing network architectures. *Annu. Rev. Genet*. **2009,** *43*, 197-222.
8. Miller, M. B.; Bassler, B. L. Quorum sensing in bacteria. *Annu. Rev. Microbiol*. **2001**, *55(1)*, 165-199.
9. Chopp, D. L.; Kirisits, M. J.; Moran, B.; Parsek, M. R. A mathematical model of quorum sensing in a growing bacterial biofilm., *J. Ind. Microbiol. Biotechnol*. **2002**, 29(6), 339-346.
10. Eberl, H. J.; Parker, D. F.; Vanloosdrecht, M. C. (2001). A new deterministic spatio-temporal continuum model for biofilm development. *Comput. Math. Methods Med*. **2001**, *3(3)*, 161-175.
11. Kreft, J. U.; Picioreanu, C.; Wimpenny, J. W.; van Loosdrecht, M. C. (2001). Individual-based modelling of biofilms. *Microbiology*, **2001**, *147(11)*, 2897-2912.
12. Dilanji, G. E.; Langebrake, J. B.; De Leenheer, P.; Hagen, S. J. (2012). Quorum activation at a distance: spatiotemporal patterns of gene regulation from diffusion of an autoinducer signal. *J. Am. Chem. Soc.,* **2012**, *134(12)*, 5618-5626.
13. Kosakowski, J.; Verma, P.; Sengupta, S.; Higgs, P. G. The evolution of antibiotic production rate in a spatial model of bacterial competition. *Plos one,* **2018**, *13(10)*, e0205202
14. Ben-Jacob, E., Schochet, O., Tenenbaum, A., Cohen, I., Czirok, A., & Vicsek, T. Generic modelling of cooperative growth patterns in bacterial colonies. *Nature*, **1994**, *368(6466)*, 46-49.
15. Goryachev, A. B.; Toh, D. J.; Wee, K. B.; Lee, T.; Zhang, H. B.; Zhang, L. H. Transition to quorum sensing in an Agrobacterium population: A stochastic model. *PLoS Comput. Biol*. **2005**, *1(4)*, e37.
16. Abadal, S.; Akyildiz, I. F. Automata modeling of quorum sensing for nanocommunication networks. *Nano Commun. Netw*, **2011**, *2(1)*, 74-83.
17. Roy, A., Goberman, D., & Pugatch, R. (2021). A unifying autocatalytic network-based framework for bacterial growth laws. *PNAS* **2021,** 118(33), e2107829118.
18. Dockery, J. D.; Keener, J. P. A mathematical model for quorum sensing in Pseudomonas aeruginosa. *Bull. Math. Biol.,* **2001**, *63(1)*, 95-116.
19. Williams, P.; Camara, M.; Hardman, A.; Swift, S.; Milton, D.; Hope, V. J.; ... Bycroft, B. W. Quorum sensing and the population-dependent control of virulence. *Philos. Trans. R. Soc. Lond.*, *B, Biol. Sci.* , **2000**, *355(1397)*, 667-680.
20. Alfinito, E.; Reggiani, L. Mechanisms responsible for the photocurrent in bacteriorhodopsin. *Phys. Rev. E*, **2015**, *91(3)*, 032702.
21. Alfinito, E.; Reggiani, L. Opsin vs opsin: New materials for biotechnological applications. *J. Appl. Phys.* **2014** *116(6)*, 064901.
22. Angert, E. R. Alternatives to binary fission in bacteria. *Nat. Rev. Microbiol.*, **2005**, *3(3)*, 214-224.
23. Alfinito, E.; Beccaria, M.; Macorini, G. Critical behavior in a stochastic model of vector mediated epidemics. *Sci. Rep.* **2016**, *6(1)*, 1-11.
24. Alfinito, E.; Beccaria, M.; Fachechi, A.; Macorini, G. Reactive immunization on complex networks. *EPL,* **2017,** *117(1)*, 18002.
25. Doh, I. J.; Kim, H.; Sturgis, J.; Rajwa, B.; Robinson, J. P.; Bae, E. Optical multi-channel interrogation instrument for bacterial colony characterization. *PloS one*, **2021**, *16(2)*, e0247721.
26. Talà, A.; Delle Side; D.; Buccolieri, G.; Tredici, S. M.; Velardi; L., Paladini; F., ... Alifano, P. Exposure to static magnetic field stimulates quorum sensing circuit in luminescent Vibrio strains of the Harveyi clade. *PLoS One*, 9(6), **2014**, e100825.
27. Borole, A. P.; Reguera, G.; Ringeisen, B.; Wang, Z. W.; Feng, Y.; Kim, B. H. Electroactive biofilms: current status and future research needs. *Energy Environ. Sci*. **2011**, *4(12)*, 4813-4834.
28. Chen, S., Jing, X.; Tang, J.; Fang, Y.; Zhou, S. Quorum sensing signals enhance the electrochemical activity and energy recovery of mixed-culture electroactive biofilms. *Biosens. Bioelectron*. **2017,** *97*, 369-376.